\documentclass[preprint,prd,tightenlines,nofootinbib,superscriptaddress]{revtex4-1}

\usepackage{color}
\usepackage{graphicx}
\definecolor{red}{rgb}{1,0,0}
\def\lesssim{\ \hbox{\raise 2pt \hbox{$<$} \kern -13pt
                     \lower 3pt \hbox{$\sim$}}\ }
\def\greatersim{\ \hbox{\raise 2pt \hbox{$>$} \kern -13pt
                     \lower 3pt \hbox{$\sim$}}\ }
\def\cascade{{\sc Cascade}}
\def\pythia{{\sc Pythia}}

\def\powheg{{\sc Powheg}}
\def\mcatnlo{{\sc Mc@nlo}}
\input epsf.tex
\def\desepsf(#1 width #2){\epsfxsize=#2 \epsfbox{#1}}

\usepackage{amsmath,bm}
\begin{document}

\hspace*{12.9 cm} {\small DESY 12-166} 

\hspace*{12.9 cm} {\small OUTP-12-19P}

\vspace*{1.4 cm} 

\title{Longitudinal momentum shifts, showering and nonperturbative   
 corrections  in matched NLO-shower  event generators}
\author{S.\ Dooling}
\affiliation{Deutsches Elektronen Synchrotron, D-22603 Hamburg}
\author{P.\ Gunnellini}
\affiliation{Deutsches Elektronen Synchrotron, D-22603 Hamburg}
\author{F.\ Hautmann} 
\affiliation{Theoretical Physics Department, 
University of Oxford,    Oxford OX1 3NP}
\affiliation{Physics \& Astronomy, 
University of   Sussex,   Brighton   BN1  9QH}
\author{H.\ Jung}
\affiliation{Deutsches Elektronen Synchrotron, D-22603 Hamburg}
\affiliation{Elementaire Deeltjes Fysica, Universiteit Antwerpen, B 2020 Antwerpen}

\begin{abstract}
   Comparisons of  experimental 
   data         with    theoretical  predictions 
    for collider processes containing  hadronic jets 
       rely on  shower  Monte Carlo event  generators   
   to include     corrections  to perturbative calculations  from hadronization, 
 parton showering, multiple parton collisions.   
We examine current  treatments   of  these 
corrections 
and propose alternative methods to 
 take into account  nonperturbative effects and  parton showering  
in the context of next-to-leading-order (NLO) event generators. 
We point out sizeable 
parton-showering corrections to jet transverse energy 
spectra at high rapidity, and discuss kinematic shifts 
in longitudinal momentum distributions from initial state showering in 
the case  both  of  jet   production and  of 
heavy mass production   at the Large 
Hadron Collider.  
\end{abstract}


\maketitle

\section{Introduction}

Phenomenological analyses of   collider processes 
involving the  production of hadronic jets 
rely  on event simulation 
by    parton shower Monte Carlo  generators~\cite{herwref,pythref}.   
The subject of this paper concerns two  different, common uses 
of  shower Monte Carlo generators:    
one in which  they are  combined    with 
   hard scattering matrix elements   
    via   a  matching  scheme, 
   e.g.  at the next-to-leading order (NLO)~\cite{ma,hoe}  in 
perturbative QCD;  another  in which  they  are  used to 
obtain   corrections   to  perturbative calculations   
  due to hadronization, showering and 
multiple parton interactions (see e.g.~\cite{atlas-1112,CMS:2011ab}),  
with such 
correction factors  then being  applied  to  determine 
realistic predictions, which can  be compared  with experimental data.

 We begin in   Sec.~II  by considering methods to  evaluate 
the  nonperturbative (NP)  corrections  to jet cross sections  using shower    
event generators. We also estimate the corrections which arise from the initial state 
and final state parton showers and  observe   that  they are sizeable (beyond NLO) in  jet transverse energy  spectra over the full range of rapidity. 
We propose a decomposition of the corrections to be applied to fixed NLO calculations, consisting of a truly NP contribution supplemented with a  contribution coming from all order resummation via parton showers.
  
  Next, in Sec.~III   we  investigate  kinematic  aspects of  parton  showers  
  associated to     combining  the 
approximation of collinear,  on-shell partons   with  energy-momentum 
 conservation.   The main effect   is 
 an event-by-event  shift in  longitudinal 
momentum distributions 
 whose size    depends  on the observable and on the 
phase space region, and 
   increases   with increasing rapidities.  
 We illustrate this  by  numerical  Monte Carlo results 
   in different phase space regions for    four  specific  examples  
   of    jet,  heavy-quark,   electroweak gauge-boson and Higgs boson  production.  
First results   on  kinematic shifts  have  been  presented      in~\cite{coki-1209}. 
   
The   approach  of  this work    may  be helpful  to  
   analyze     corrections to finite-order 
perturbative calculations  for jet observables 
from parton showering and 
   nonperturbative   dynamics.  
  These encompass both final state fragmentation  
 effects and initial state   contributions associated with 
  collinearity approximations.  
Dynamical  high-energy effects  
   on jet final states,   distinct from the ones 
  discussed in this paper,   have been 
    emphasized in~\cite{mw92,hj_ang,jhep09}    
due to    noncollinear     contributions to    parton  branching  processes. 
 We note that both these results  and the results in this paper  
 stress   the  phenomenological relevance of  
   more complete 
    descriptions of QCD parton cascades  
    in terms  of transverse momentum dependent parton 
   fragmentation  and parton density 
    functions~\cite{jccbook,avsar11,unint09,muld-rog-rev}. 
 Concluding      comments   on the results of this  work 
  are given in  Sec.~IV.

\section{Monte Carlo nonperturbative correction factors}

In this section we consider   methods to  evaluate 
 NP and parton shower correction factors.  To be definite, we refer  
 to   the  case of  inclusive production of single jets  
at the  LHC~\cite{klaus-r}.  In order to 
compare  theory with experimental  
data corrected to stable particle level, Refs.~\cite{atlas-1112,CMS:2011ab}   
supplement  NLO perturbative calculations  with  NP corrections 
% taking into account hadronization ({\em had}), multiple 
%parton interactions ({\em mpi}),  parton showering ({\em ps}). These contributions   are 
  estimated from Monte Carlo  event  generators. 
 Using leading-order  Monte Carlo (LO-MC) generators~\cite{herwref,pythref}, the 
 correction factors $K_0$ are schematically obtained by~\cite{atlas-1112,CMS:2011ab} 
\begin{equation} 
\label{npK1} 
K^{NP}_{0} =  { {  N_{LO-MC}^{(ps+mpi+had)} } /  {  N_{LO-MC}^{(ps)} }}  \;\; ,
\end{equation} 
where $(ps+mpi+had)$  and   $(ps)$ mean 
respectively  a simulation including   
parton showers, multiparton interactions and hadronization,  
and   a simulation including  only 
 parton showers     in addition to the LO hard process. 
Having only LO+PS event generators available,  this  is the most obvious way to 
estimate NP   corrections to be applied to NLO parton level calculations.
However, when these  corrections  are combined   
 with NLO parton-level results, a potential 
   inconsistency arises  because the 
    radiative correction  from the first  gluon emission   is treated  at different levels  of 
    accuracy     in the two parts of the  calculation.        

We here suggest that  an alternative method which    avoids this  is to use 
NLO Monte Carlo (NLO-MC) generators to determine the correction. 
In this case one can consistently  assign correction factors 
to be applied to  NLO calculations.  Moreover, this  method  allows one to 
study  separately 
 correction factors to the  fixed-order calculation  due to  parton showering  
effects. 
To  this end, we  introduce 
  the correction factors $K^{NP}$  and  $K^{PS}$ as 
\begin{equation} 
\label{npK2} 
K^{NP} =  { {  N_{NLO-MC}^{(ps+mpi+had)} } /  {  N_{NLO-MC}^{(ps)} }}   \;\;   , 
\end{equation} 
\begin{equation} 
\label{npK3} 
K^{PS} =  { {  N_{NLO-MC}^{(ps)} } /  {  N_{NLO-MC}^{(0)} }}   \;\;   ,  
\end{equation}
where    the  denominator  in Eq.~(\ref{npK3})   is  defined  by switching off 
all components beyond NLO in the Monte Carlo simulation. 
The difference  between  the correction factors in   Eqs.~(\ref{npK1}) and (\ref{npK2}) 
 comes primarily  from the way in which  the multiple parton interaction (MPI) contribution 
   is  matched to the NLO calculation.  MPI processes have typical transverse momentum  
   scales smaller than the scale   of the hard process, which may  be defined as the average transverse  momentum of the hard partons. This however is different in 
    LO and NLO calculations, giving 
   rise to non-negligible   numerical  differences, which we will  show below.    
The correction factor in 
Eq.~(\ref{npK3}), on the other hand, is new.  It  singles out contributions 
due to parton showering.  This correction factor  has not been   considered in 
earlier analyses.  
We    show  below  its  numerical significance.   We anticipate 
that  taking properly  into account   these  showering corrections    
  can  be     relevant     in    fits  for 
parton distribution functions   using  inclusive jet data.

\begin{figure}[htbp]
\begin{center}
\includegraphics[scale=.4]{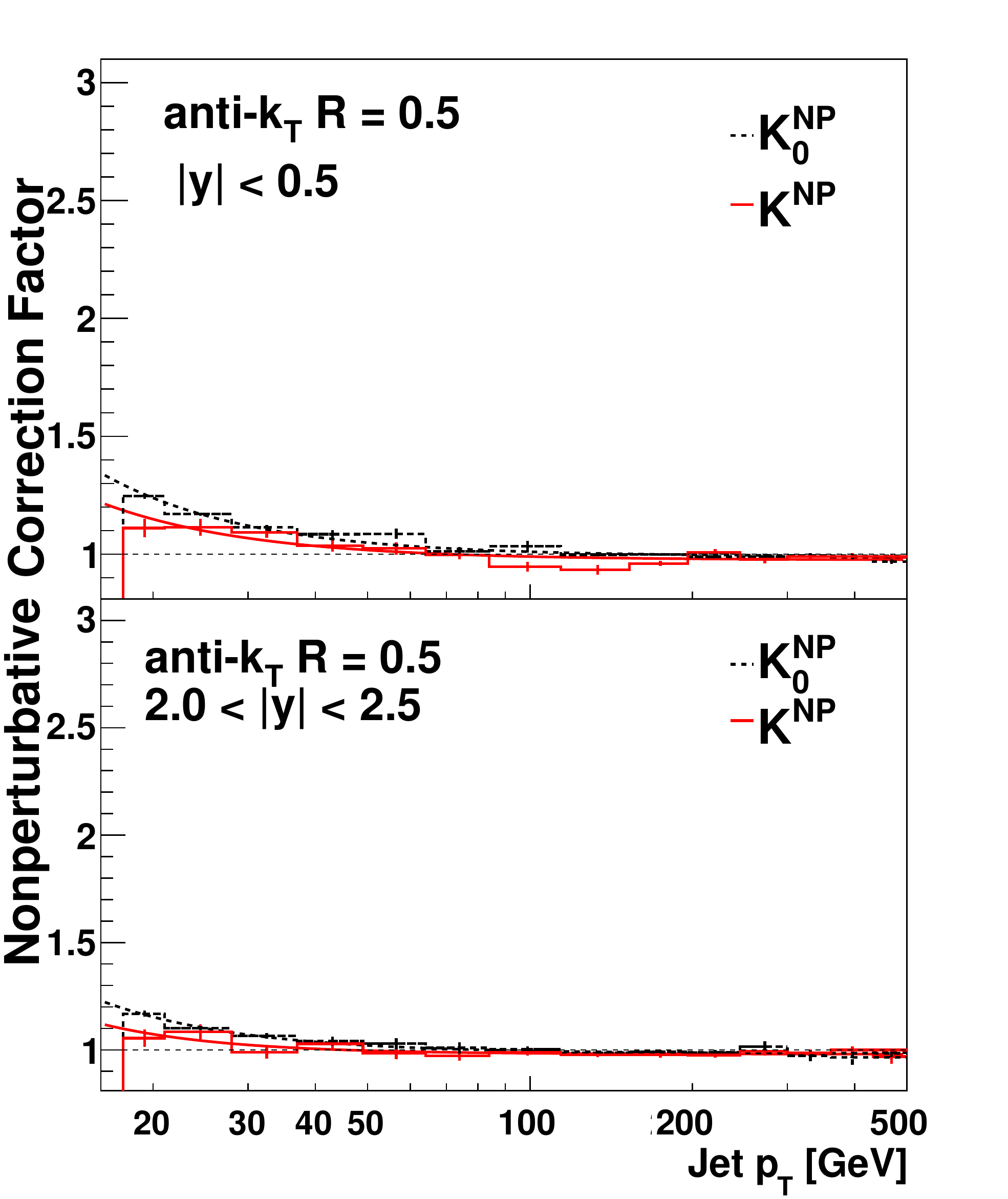}
\includegraphics[scale=.4]{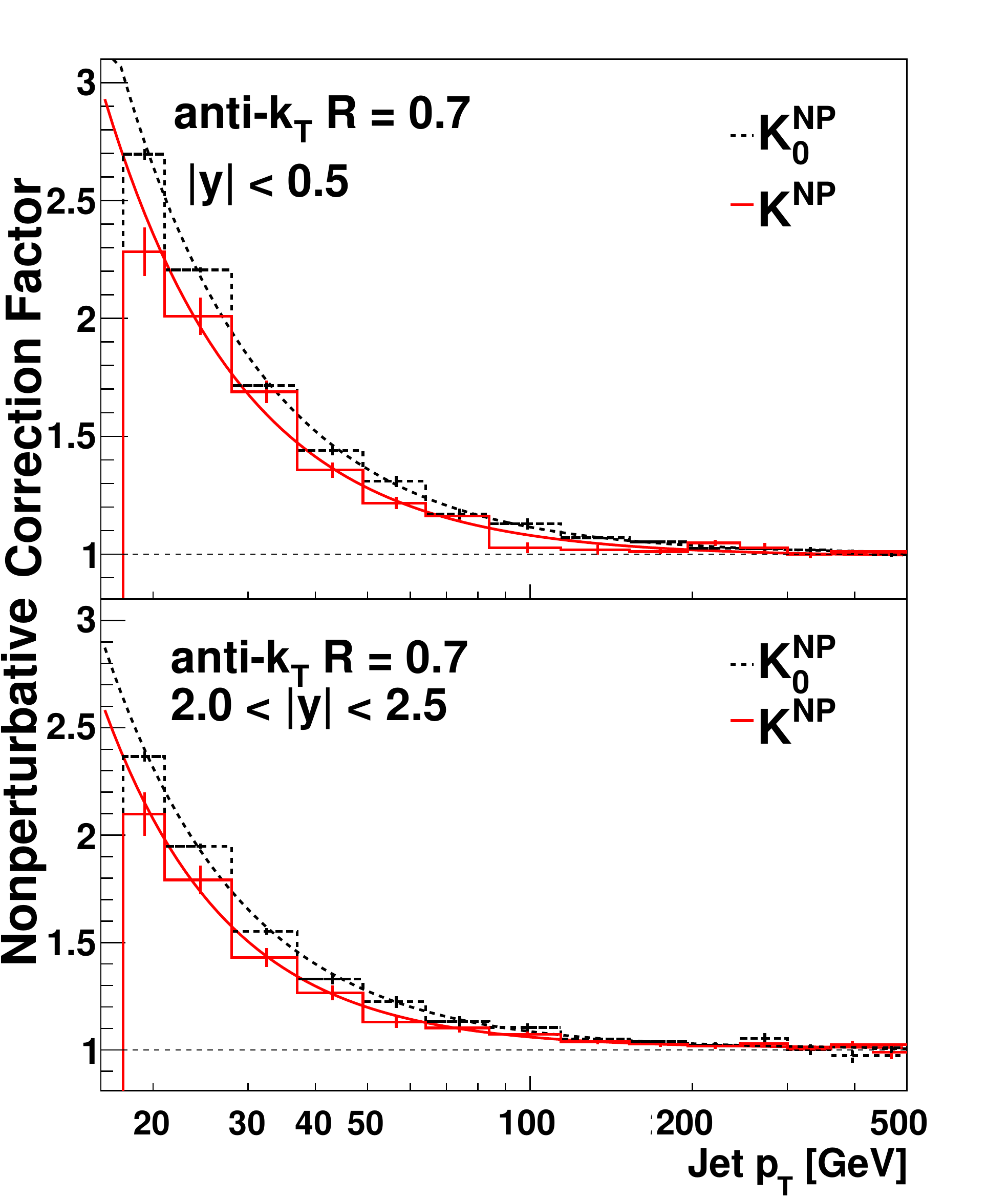}
\caption{\it The NP  correction factors to jet transverse 
momentum distributions  obtained from   Eq.~(\ref{npK1}) and 
 Eq.~(\ref{npK2}),  using 
 \protect\pythia\  and \protect\powheg\  respectively, 
 for $|y|<0.5$ and $2 < |y|< 2.5$.  
Left: $R=0.5$; Right:  $R=0.7$.}  
\label{fig:np1}
\end{center}
\end{figure}

In Fig.~\ref{fig:np1}   we  compute  results   for  
the NP correction factors in Eqs.~(\ref{npK1}),(\ref{npK2})    
to jet transverse  momentum  distributions.  We define jets using the anti-k$_T$  
algorithm~\cite{antiktalgo} with  jet size $R =  0.5$  and  $R = 0.7$.  
We plot the results versus the jet transverse momentum $p_T$ for  different 
regions in  the jet rapidity $y$. 
We  show   $K^{NP}$ as obtained using 
the NLO event generator 
\powheg~\cite{alioli}  and compare it 
    to the result 
obtained  at leading order  from  \pythia~\cite{pythref} 
(tune Z2~\cite{tunes} 
and CTEQ6L1 pdfs~\cite{pumplin02}).
%This uses  the \pythia\  parton shower tune Z2~\cite{tunes} 
%and CTEQ6L1 pdfs~\cite{pumplin02}.   
The curves    in Fig.~\ref{fig:np1}  
illustrate the differences coming from the definition of the hard process.

\begin{figure}[htbp]
\begin{center}
\includegraphics[scale=.4]{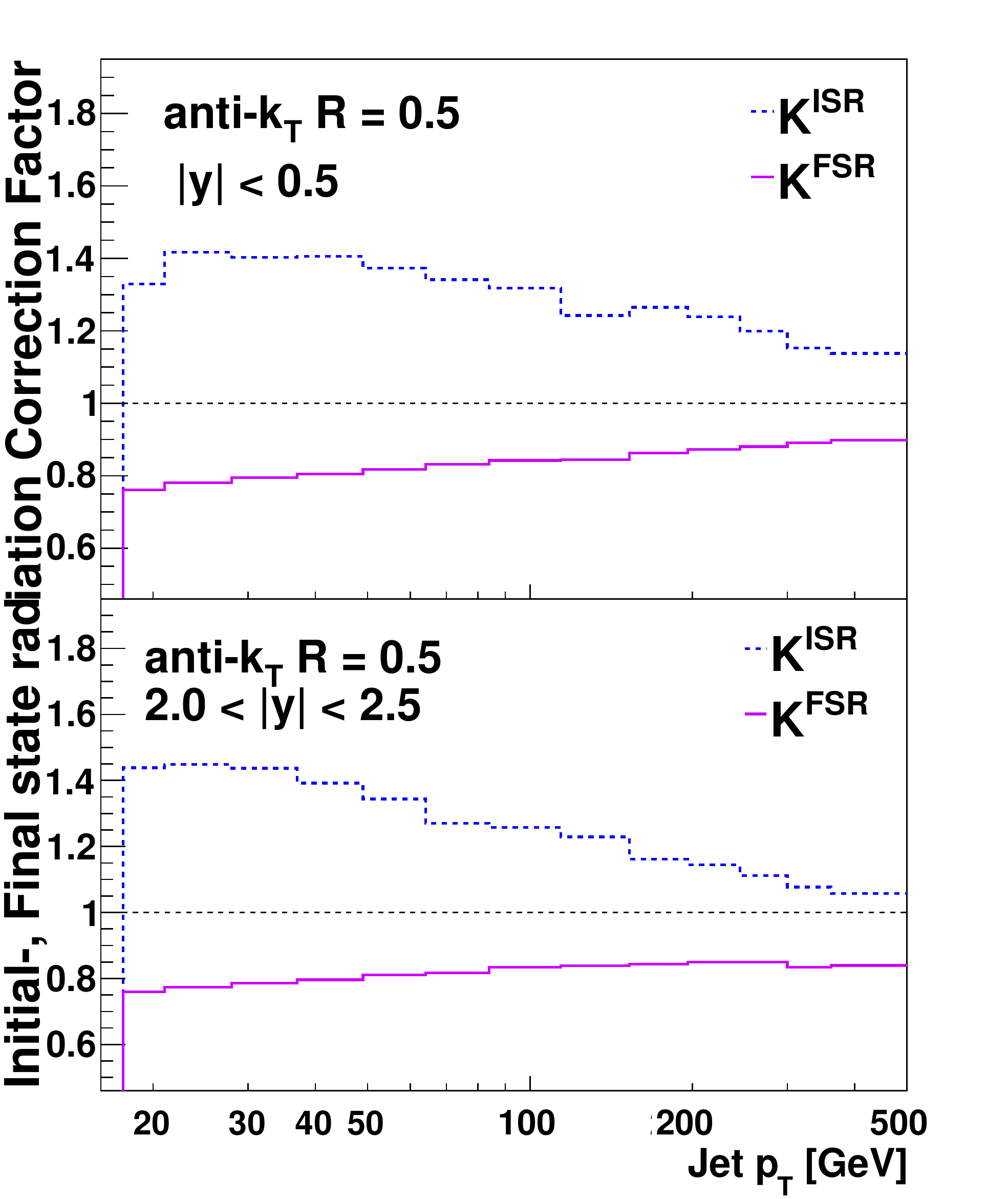}
\includegraphics[scale=.4]{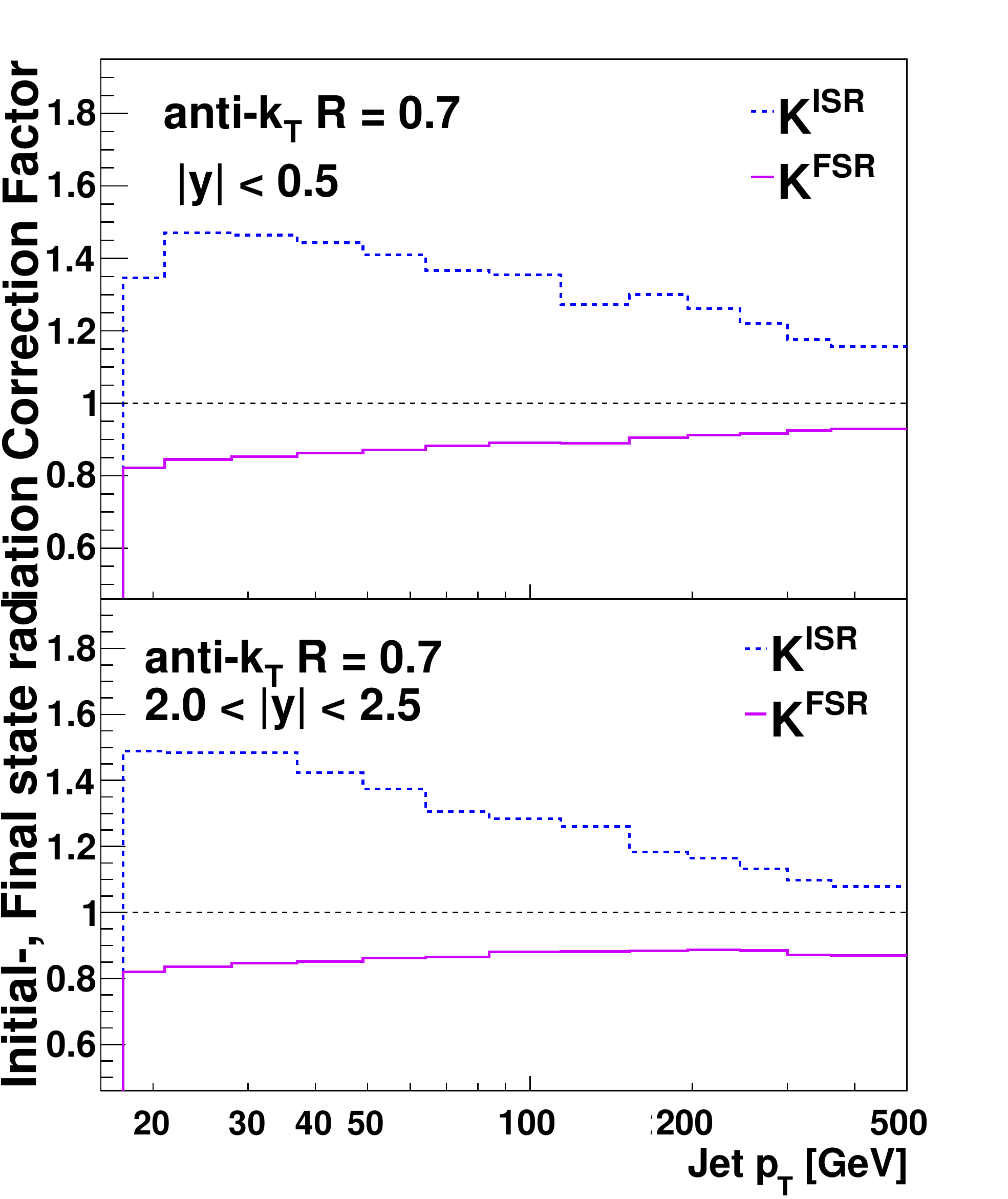}
\caption{\it The initial and final state parton shower 
 correction factor  to jet transverse momentum  distributions, obtained from  
 Eq.~(\ref{npK3}) using \protect\powheg\   
 for $|y|<0.5$ and $2 < |y|< 2.5$.
Left: $R=0.5$; Right:  $R=0.7$.}  
\label{fig:np2-IFPS}
\end{center}
\end{figure}

\begin{figure}[htbp]
\begin{center}
\includegraphics[scale=.4]{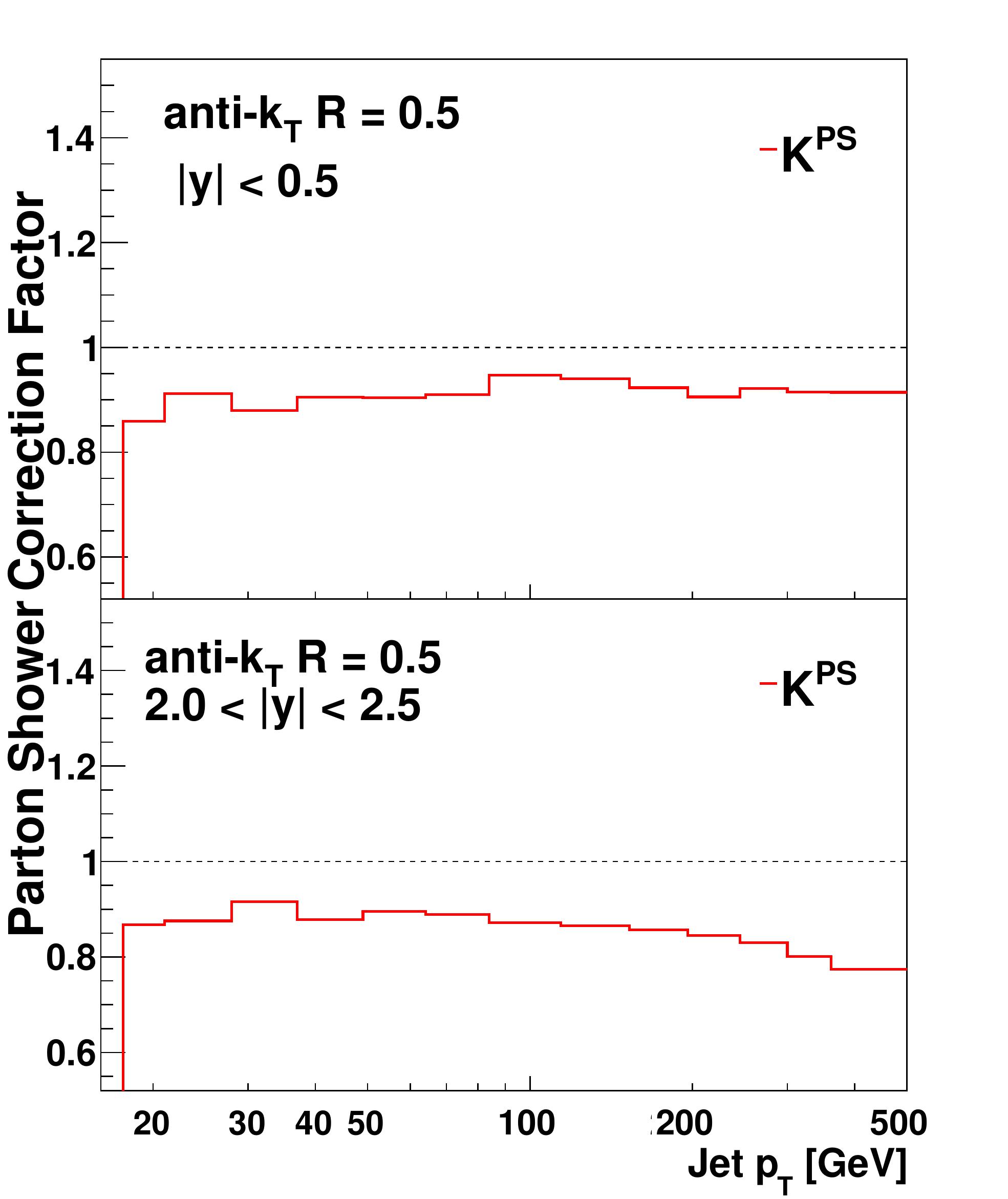}
\includegraphics[scale=.4]{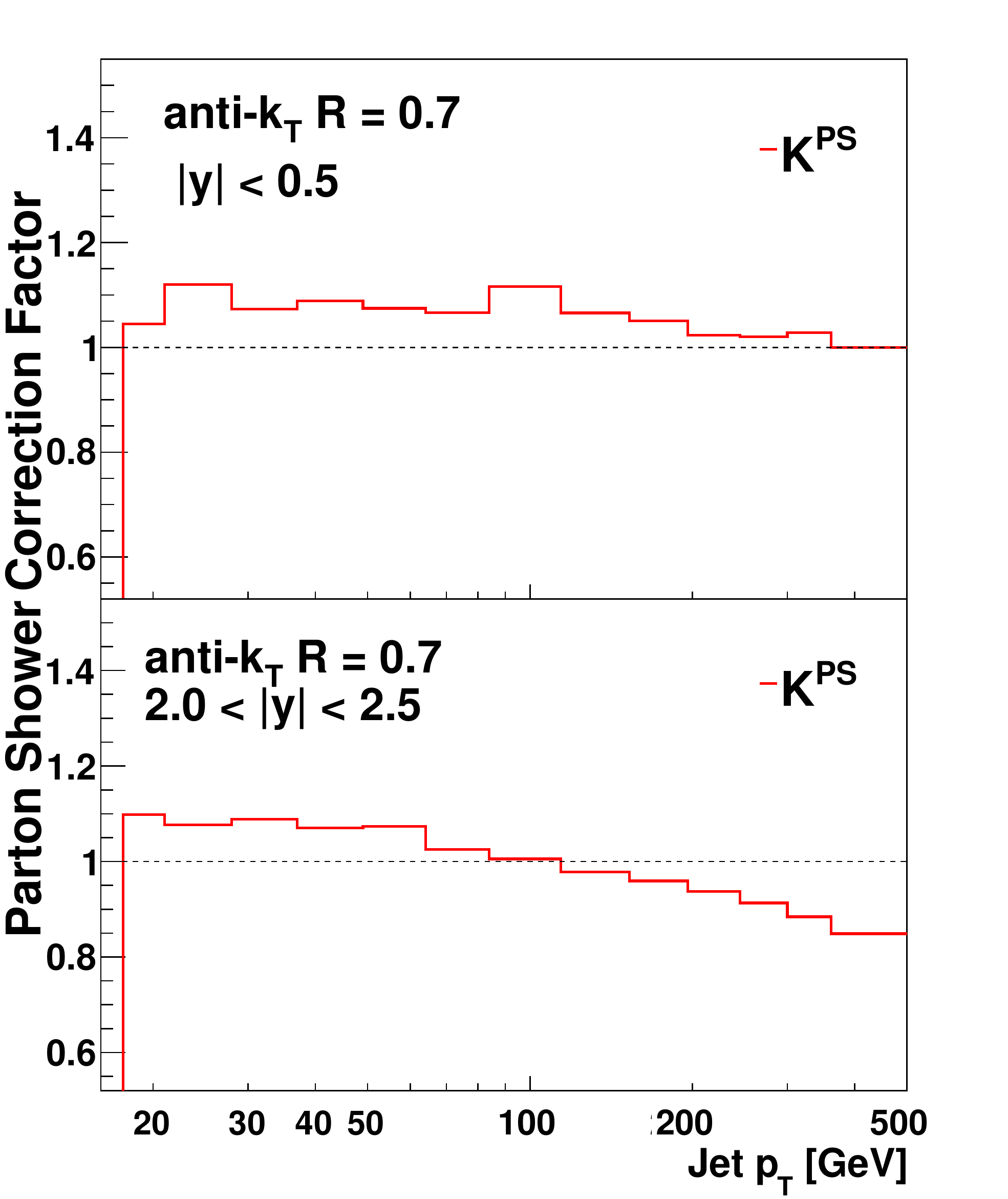}
\caption{\it The  parton shower
 correction factor to jet transverse momentum  distributions, obtained from  
 Eq.~(\ref{npK3})  using 
 \protect\powheg\   
   for $|y|<0.5$ and $2 < |y|< 2.5$.
Left: $R=0.5$; Right:  $R=0.7$.}  
\label{fig:np2}
\end{center}
\end{figure}

In Figs.~\ref{fig:np2-IFPS}   and  \ref{fig:np2} we  compute 
   the corrections  from parton shower $K^{PS}$ as obtained from  Eq.~(\ref{npK3}) 
   as a function of the jet $p_T$  
   for different values of $R$ and different  rapidities $y$.   
    Fig.~\ref{fig:np2-IFPS}  
    shows the contributions 
     coming from initial state and final state parton shower separately. 
     We note that the initial and final state showers are so interconnected 
      that the      combined effect is nontrivial and cannot be obtained 
     by simply adding the two results. 
     In general the effect from parton shower is largest at large $|y|$, where  the 
initial state parton shower is mainly contributing 
at low $p_T$, while the final state parton shower is contributing significantly 
 over the whole $p_T$ range.    
 In particular note  in Fig.~\ref{fig:np2}  that,  while 
 at  central rapidity     the combined 
 shower correction is rather flat in  $p_T$, at higher rapidity 
 this  is no longer flat and  for large 
 $p_T$   it  may  even dip  below the  correction from purely 
 final state shower  reported  in     
 Fig.~\ref{fig:np2-IFPS}. This    suggests  
  that migration effects become relevant  not only 
 in $p_T$  but also in $y$.

 While the NP corrections  studied in Fig.~\ref{fig:np1}   
 become vanishingly small  at  sufficiently  large   $p_T$, the 
 showering correction    in  Figs.~\ref{fig:np2-IFPS}   and  \ref{fig:np2}  
 gives  finite effects also for large    $p_T$.  
 %if    rapidities  are   non-central.   
Since, as shown by  our results,  the size  of this    effect  
does depend  on the value of rapidity $y$, this 
   will    influence   the 
  shape of jet distributions   and the  comparisons  of theory  predictions 
   with      experimental data.  In particular, 
       if   the    showering correction  factor is not consistently  taken 
   into account,       besides the 
    NP corrections,  this  may    affect the determination of parton distribution  functions 
    from data sets   including jets.

 Note  that 
in~\cite{atlas-1112,CMS:2011ab}    
 NP correction factors   $K_0$ are  applied to  the 
  NLO calculation~\cite{nagy}, and the data comparison shows that 
 the NLO 
calculation agrees with data at central rapidities, while increasing deviations are 
seen with increasing rapidity at large transverse momentum $p_T$~\cite{atlas-1112}.   
A second  comparison is   performed  in~\cite{atlas-1112}  with    NLO-matched 
\powheg\  calculations~\cite{alioli}, showing   
 large differences  
in the high  rapidity region between   results obtained by  interfacing 
\powheg\   with different shower  models~\cite{herwref,pythref}  and 
different model tunes~\cite{tunes-sk,tunes}.~\footnote{Further  discussion 
 of   parton  showering effects  on high-rapidity jets  may be found  in~\cite{fwd-phen}.}   
Motivated by this observation,  in the next section we consider more closely the 
kinematics of the initial state parton shower at high rapidity.

\section{Initial state showering and  kinematic shifts}

\begin{figure}[htb]
%\vspace{45mm}
%\special{psfile=fig_sec2_rev.eps hscale=36 vscale=36
%  hoffset=175 voffset=5}
\includegraphics[scale=.5]{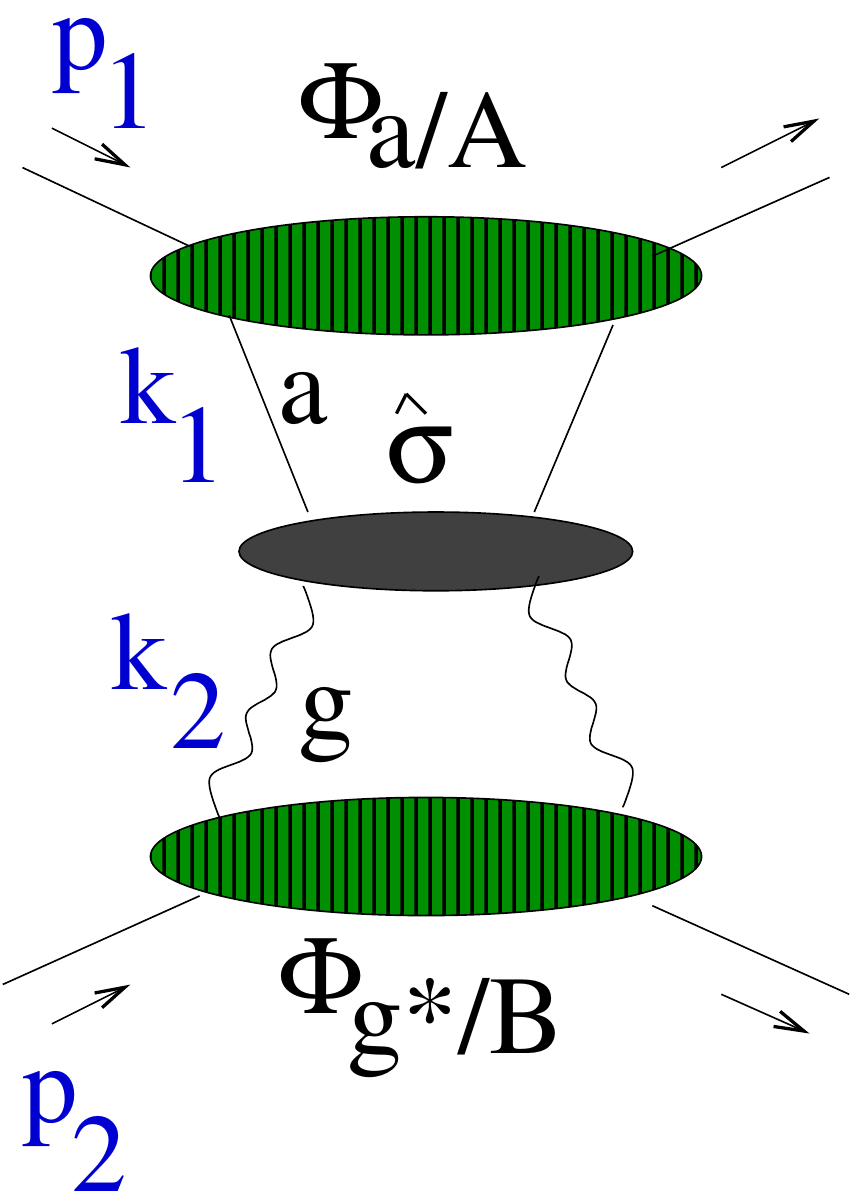}
\caption{\it Factorized structure of the jet cross section at high rapidity.} 
\label{fig:sec2}
\end{figure}

Let us    recall 
the physical picture~\cite{jhep09} of  jet production at high rapidity     (Fig.~\ref{fig:sec2})   
based on 
QCD high-energy factorization~\cite{hef}. 
Take  the incoming momenta $p_1$ and $p_2$ in 
Fig.~\ref{fig:sec2}   
  in the plus and minus  lightcone 
directions, 
defined, for any four-vector $v^\mu$, as $v^\pm = (v^0 \pm v^3 ) /  \sqrt{2}$. 
Let us  
parameterize the exchanged momenta $k_1$ and $k_2$  in terms of     
purely transverse four-vectors $k_{\perp 1 }$  and $k_{\perp 2}$ and 
longitudinal (lightcone) 
momentum  fractions $x_i$  (collinear)  and $ {\overline x}_i$    (anti-collinear) as  
$ 
%\begin{equation}
%\label{kinek1k2}
   k_1 =   x_1  p_1 + k_{\perp 1} + {\overline x}_1 p_2  
$,  $ 
%   \;\;, \;\;\; 
   k_2 =  x_2  p_2 + k_{\perp 2} + {\overline x}_2 p_1 
%   \;\; . 
%\end{equation}
$.  
To single-logarithmic accuracy in the jet rapidity and the jet  transverse momentum, 
we  may  approximate  
$k_1$ and $k_2$  
 using strong ordering in  the longitudinal momenta, and  get~\cite{jhep09}   
\begin{equation}
\label{fwdkin}
k_1 \simeq   x_1  p_1 \;\;\;   ,  \;\;\;\;\;  k_2 \simeq   x_2  p_2  +  k_{\perp 2}   
 \;\;\;   ,  \;\;\;\;\;  
x_1 \gg x_2     \;\;  .  
\end{equation}
The physical picture    corresponding to the   factorization~\cite{jhep09,hef}   
consists of the scattering  of   a highly off-shell, low-$x$ parton  off a nearly on-shell, 
high-$x$ parton. The calculations~\cite{jhep09,fwd-phen} embody this picture through the 
longitudinal and  transverse momentum dependences of both perturbative and 
nonperturbative components of the jet  cross section, denoted respectively by  ${\widehat \sigma}$ 
and $\Phi$ in Fig.~\ref{fig:sec2}.  In what follows, however,  we 
will not use the   specific    content 
of these calculations, but  we will  simply  use the   
underlying  physical   picture as a guidance 
  to examine  kinematic  effects   of  collinear  approximations.

In the light of this picture,  let us   consider the 
NLO-matched   shower  Monte Carlo calculations, following~\cite{coki-1209}. In the 
 Monte Carlo event generator first  the
 hard subprocess events with full four-momentum assignments 
for the external lines are generated.  In particular,  the momenta $k_j^{(0)}$  ($ j = 1 , 2 $) of the 
partons initiating the hard scatter are on shell, and are taken to be fully collinear 
with the incoming  state  momenta $p_j$, 
\begin{equation}
\label{k-before-sho} 
 k_j^{(0)} = x_j p_j  \;\;\;\;\;\;\;\;   ( j = 1 , 2 ) \;\;\; . 
\end{equation}  
  Next  the 
showering algorithm is applied, and complete final states are generated including 
additional QCD radiation from the initial state and final state parton cascades. 
As a result of QCD  showering, the momenta $k_j$  are no longer exactly  collinear,  
\begin{equation}
\label{k-after-sho} 
 k_j  \neq  x_j p_j  \;\;\;\;\;\;\;\;   ( j = 1 , 2 ) \;\;\; . 
\end{equation}   
Their transverse momentum is   to be   compensated  by a change  in the 
kinematics of the hard scattering subprocess. 
By  energy-momentum conservation, however,   this  implies  a 
reshuffling, event by event,  in  the longitudinal momentum 
fractions $x_j$ of the partons  scattering off each other in the hard  subprocess. 
The size of the shift in  $x_j$ depends on the emitted transverse momenta. 

\begin{figure}[htbp]
\includegraphics[scale=.8]{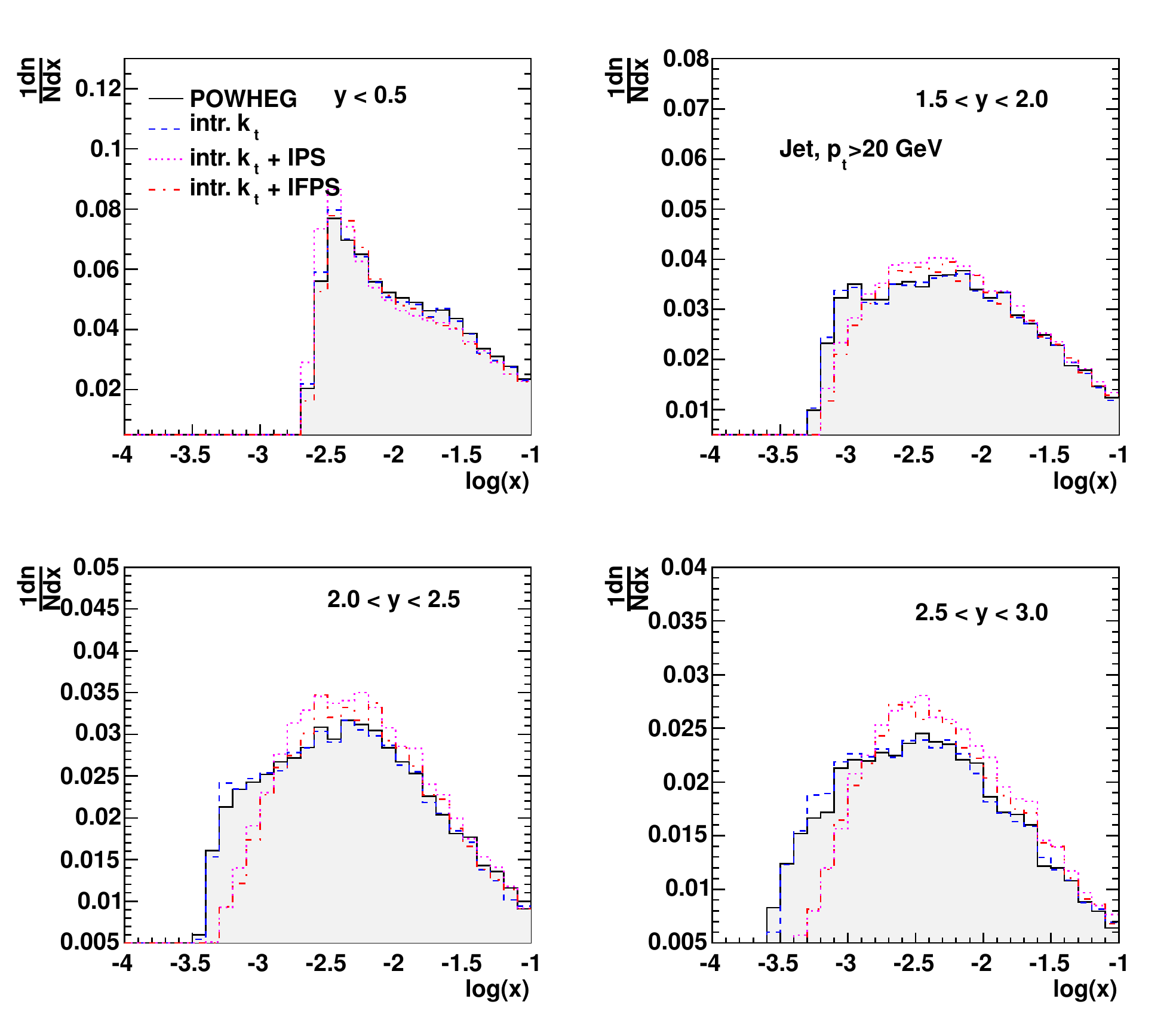}
\caption{\it Distributions in the parton longitudinal momentum fraction $x$  before (POWHEG) and after parton showering (POWHEG+PS),  
for  inclusive jet  production  at  different rapidities for jets with $p_T> 18 $ GeV obtained 
by  the anti-kt jet  algorithm~\cite{antiktalgo} with $R=0.5$.
Shown is the effect of intrinsic $k_t$, initial (IPS) and initial+final state (IFPS) parton shower.}  
\label{fig:fig1}
\end{figure}

Let us  now focus  on  jets  measured in the rapidity  
 range $ y < 2.5$~\cite{CMS:2011ab} and  examine the effect  of 
  the kinematical shift  in 
the longitudinal momentum fractions.  To this end we  compute   the distribution 
in $x_j$  from   \powheg\  before  parton showering 
and after parton  showering~\cite{coki-1209}. 
   Fig.~\ref{fig:fig1}  shows  the   distribution %for the  lower 
   for one of the $x_j$ partons. % in the initial  state     corresponding to different rapidity  intervals.  
   We   plot the result   
before showering (\powheg)  and  the results of 
successively  
including  intrinsic $k_t$, initial state parton shower  and initial+final state  parton shower. 
The results are obtained using the \pythia\  parton shower (tune Z2~\cite{tunes} 
and CTEQ6L1 pdfs~\cite{pumplin02}). This   does  not  include  
multiple parton interaction and hadronization effects.
Using  the definition of lightcone momentum fractions  
given at the beginning of this section, 
the kinematic variable $x$ is computed as $x=(E+p^z)/(2E_{\rm{beam}})$, where 
 $E$ and $p^z$ are the energy and $z$-component of momentum of 
 parton $j$, and $E_{\rm{beam}}$ is the energy of the hadron beam. 
The momentum fraction   $x$ is first calculated for the partons given by 
\powheg\ before shower and then calculated from the 
 \pythia\   event record after shower.

We see   from   Fig.~\ref{fig:fig1}   that the  kinematical  reshuffling in the 
longitudinal momentum fraction    is negligible for  central rapidities but becomes 
significant for $ y >1.5$.    This  effect  characterizes the highly asymmetric 
parton kinematics,  which   becomes important  for the first time 
at the LHC  in significant regions of phase space~\cite{jhep09}. 
Since    the perturbative weight for each event is determined 
by the initial \powheg\  simulation,   predictions   of matched  NLO-shower    
calculations  for observables 
sensitive to this asymmetric region   can be affected significantly by the 
kinematical shift as shown in   Fig.~\ref{fig:fig1}. Similarly, since  the momentum reshuffling is 
done after  the  evaluation of the parton distribution functions, the kinematical shift can 
affect predictions also through the pdfs. 
It will  be of interest to examine the impact of this    phase space  
 region    on   total cross sections as well. 

\begin{figure}[htbp]
\includegraphics[scale=.8]{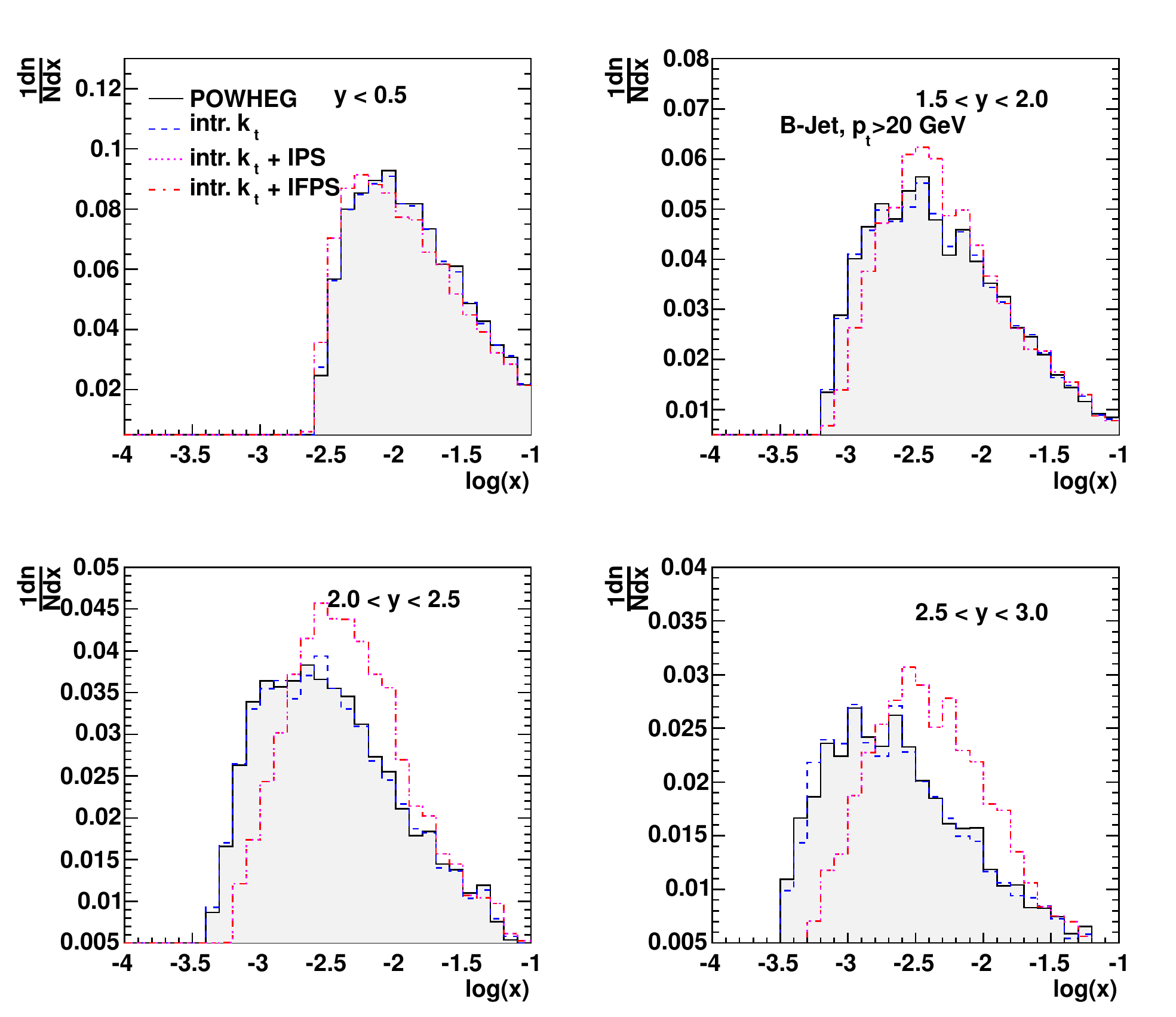}
\caption{\it  Production of $b$-jets: distribution in the parton 
longitudinal  momentum fraction $x$,  
before and after parton  showering,  for different rapidity regions.
Shown is the effect of intrinsic $k_t$, initial (IPS) and initial+final state (IFPS) parton shower.} 
\label{fig:fig2}
\end{figure} 

Let us next  consider  
 the case of bottom-flavor  jet  production~\cite{cms-bjet,atlas-bjet}. 
The   LHC measurements~\cite{cms-bjet,atlas-bjet} 
are reasonably described by NLO-matched 
shower generators  \mcatnlo~\cite{frix}  and \powheg~\cite{frix-pow}  
 at central rapidities, and 
 they are below these predictions at large rapidity and large $p_T$. 
In  Fig.~\ref{fig:fig2}  we   consider 
   $B$-jets in different rapidity regions~\cite{cms-bjet} and 
plot the  gluon $x$  distribution  
 from   \powheg\  before  parton showering and after including various 
 components of  the parton  shower generator, similarly  to what is done above for 
Fig.~\ref{fig:fig1}.  We use  the \pythia\  parton shower (tune Z2~\cite{tunes},   
here including hadronization to identify the B-jet). 
 We observe similar shift   in longitudinal momentum 
 with increasing  rapidity    as in the inclusive  jet case.

\begin{figure}[htbp]
\includegraphics[scale=.75]{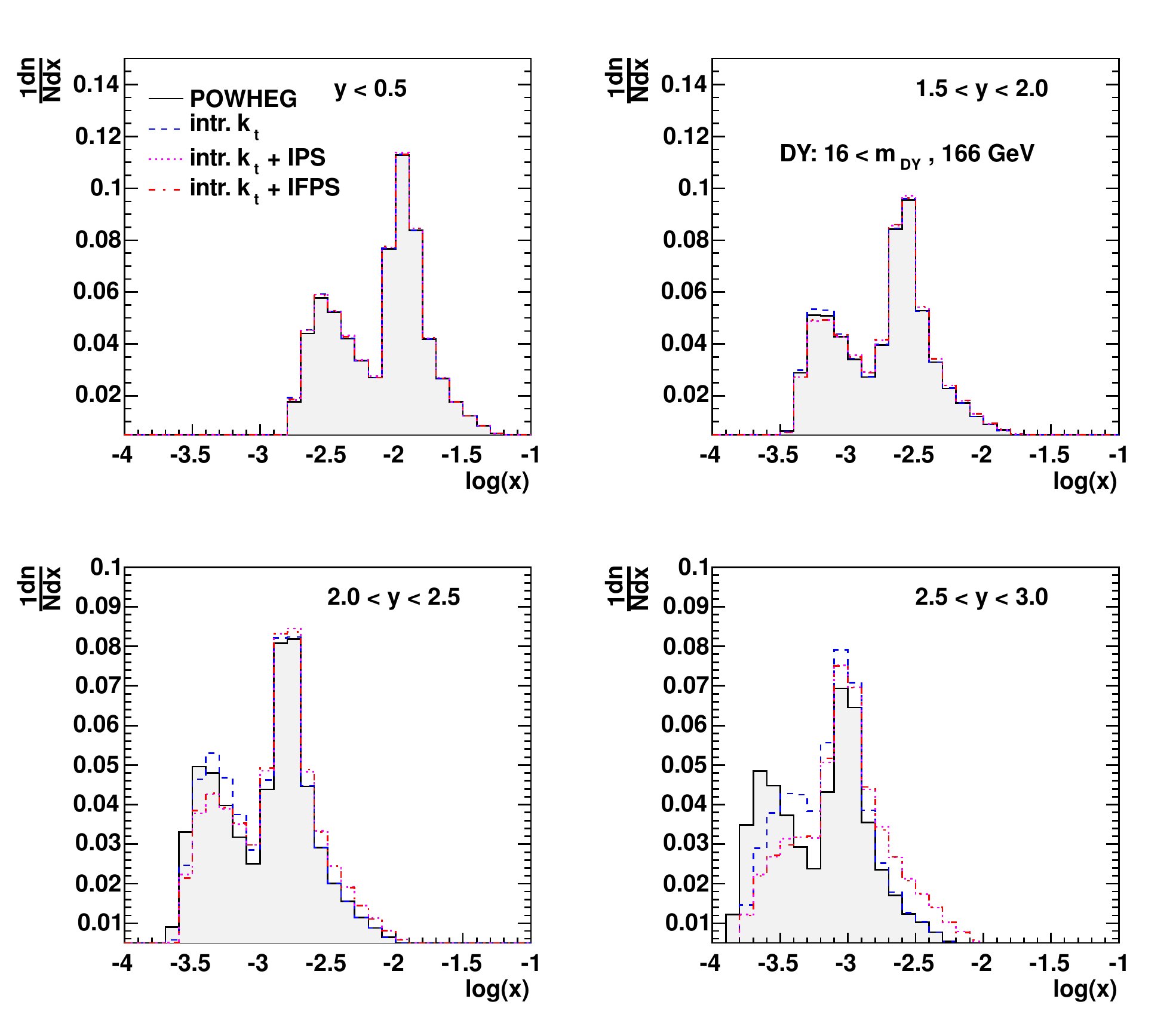}
\caption{\it  Drell-Yan production with $16 < m_{DY} < 166$ GeV: distribution in the parton 
longitudinal  momentum fraction $x$ 
 before and after showering. Shown is the effect of intrinsic $k_t$, initial (IPS) and initial+final state (IFPS) parton shower.} 
\label{fig:fig3}
\end{figure}

In  Fig.~\ref{fig:fig3}  we  consider   Drell-Yan (DY) production in the mass 
 range $16 < m_{DY} < 166$~GeV  and perform  a similar   study to what is done 
 above for jets. 
  In this case too we find that the effects of the 
kinematical reshuffling  in $x$ evaluated from    \powheg\   become   
non-negligible away from  the central rapidity region. 
The double peak structure  in  Fig.~\ref{fig:fig3}  comes from the continuum DY production in addition to  $Z_0$ production.
   It will be of   interest   to  investigate the   kinematic  reshuffling   effect  
    along with   the 
   forward Drell-Yan enhancements  discussed in~\cite{martin}.

\begin{figure}[htbp]
\includegraphics[scale=.7]{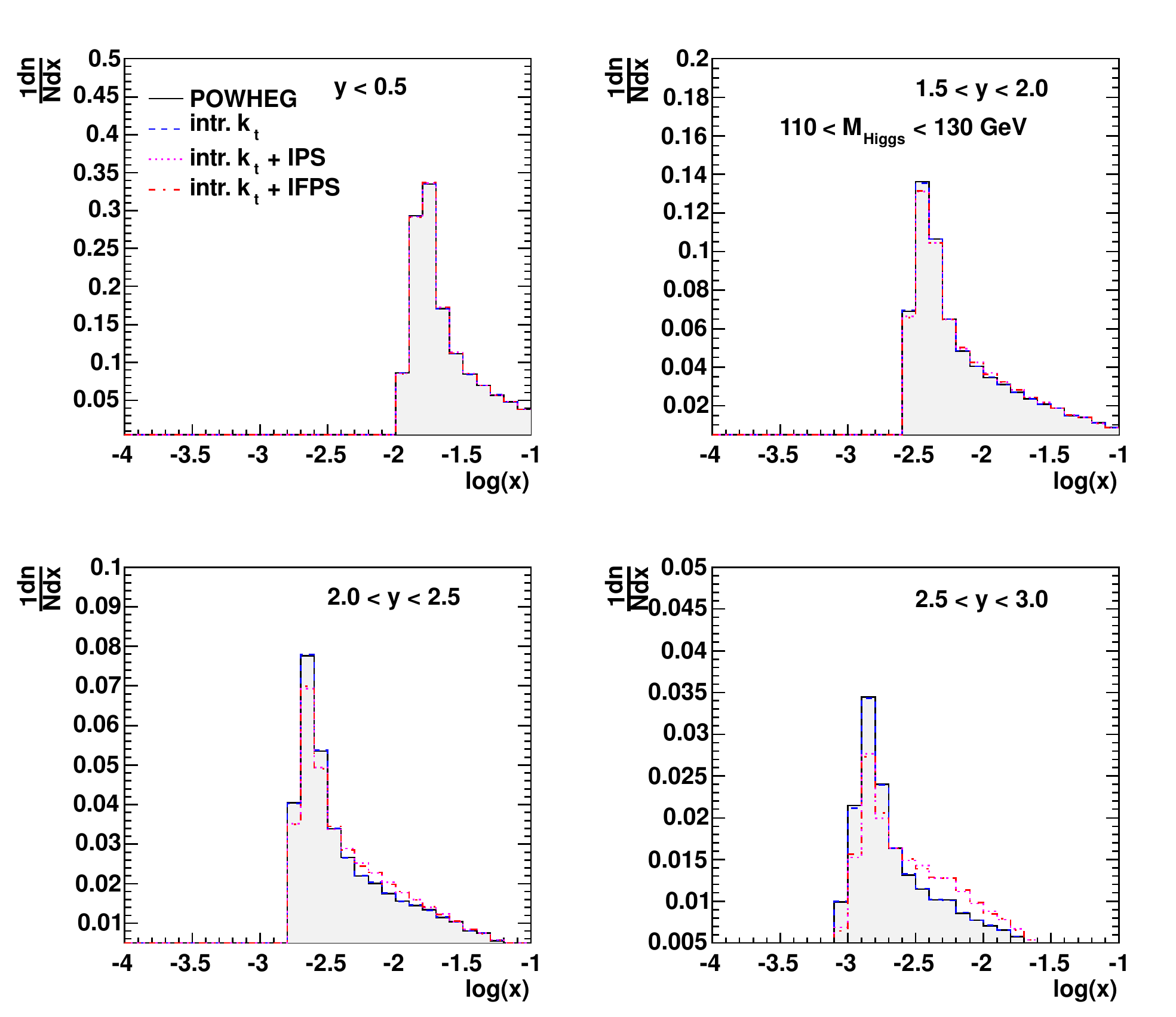}
\caption{\it  Higgs boson production with $110 < m_{Higgs} < 130$ GeV: distribution in the parton 
longitudinal  momentum fraction $x$ 
 before and after showering. Shown is the effect of intrinsic $k_t$, initial (IPS) and initial+final state (IFPS) parton shower.} 
\label{fig:fig4}
\end{figure}

Finally we  consider Higgs boson production in  
 Fig.~\ref{fig:fig4} 
 for  $110 < m_{Higgs} < 130$~GeV.    
We  observe a smaller effect at $\sqrt{s}=7$~GeV  than in the  previous cases  
since the $x$-range is  limited  by   the  
 Higgs mass. 
 
\begin{figure}[htbp]
\includegraphics[scale=.7]{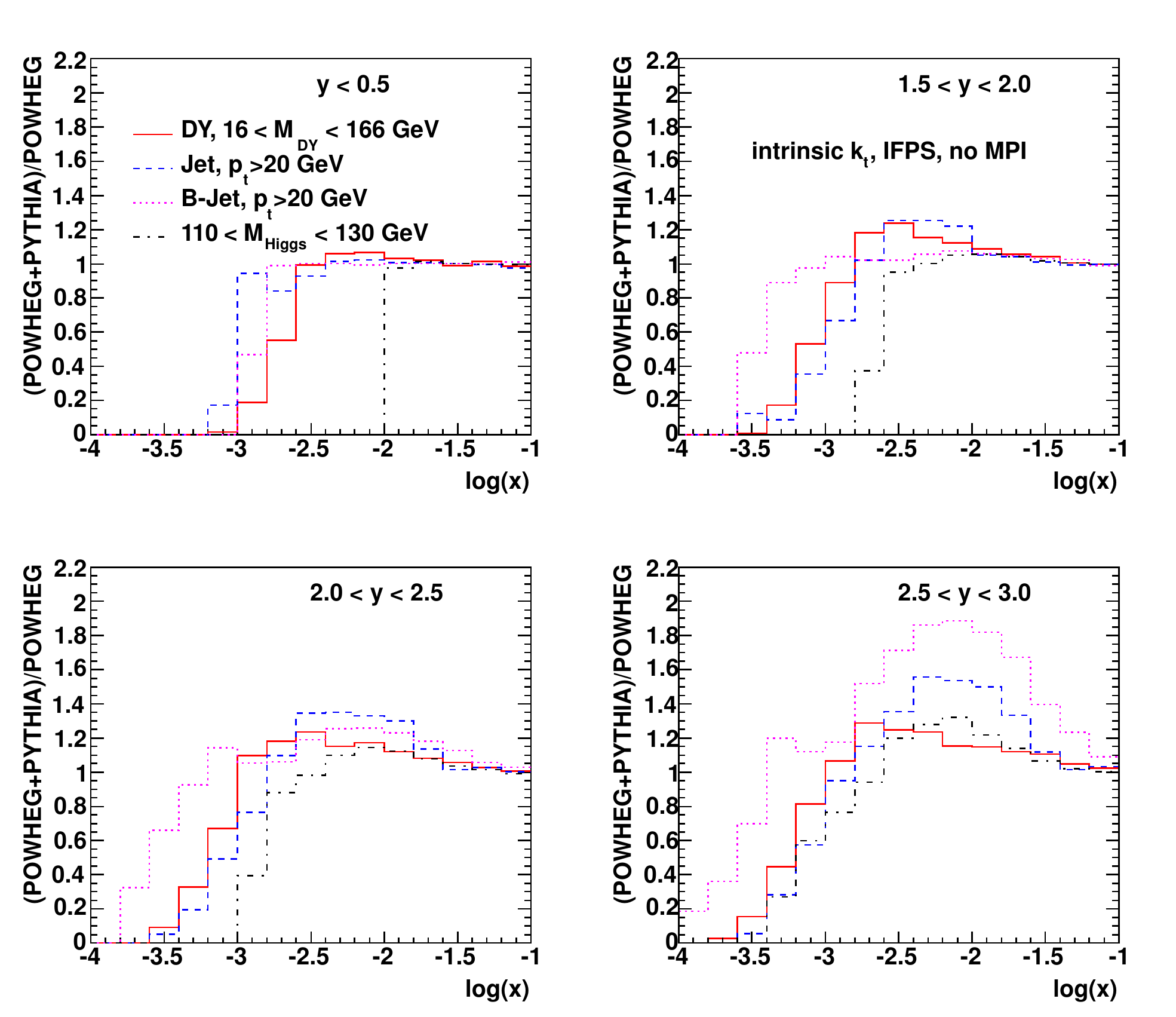}
\caption{\it  Ratio of the cross sections obtained with \protect\powheg\   after  and 
before  inclusion of initial + final state parton shower and intrinsic $k_t$ for the different processes. 
} 
\label{fig:fig5}
\end{figure}
\begin{figure}[htbp]
\includegraphics[scale=.75]{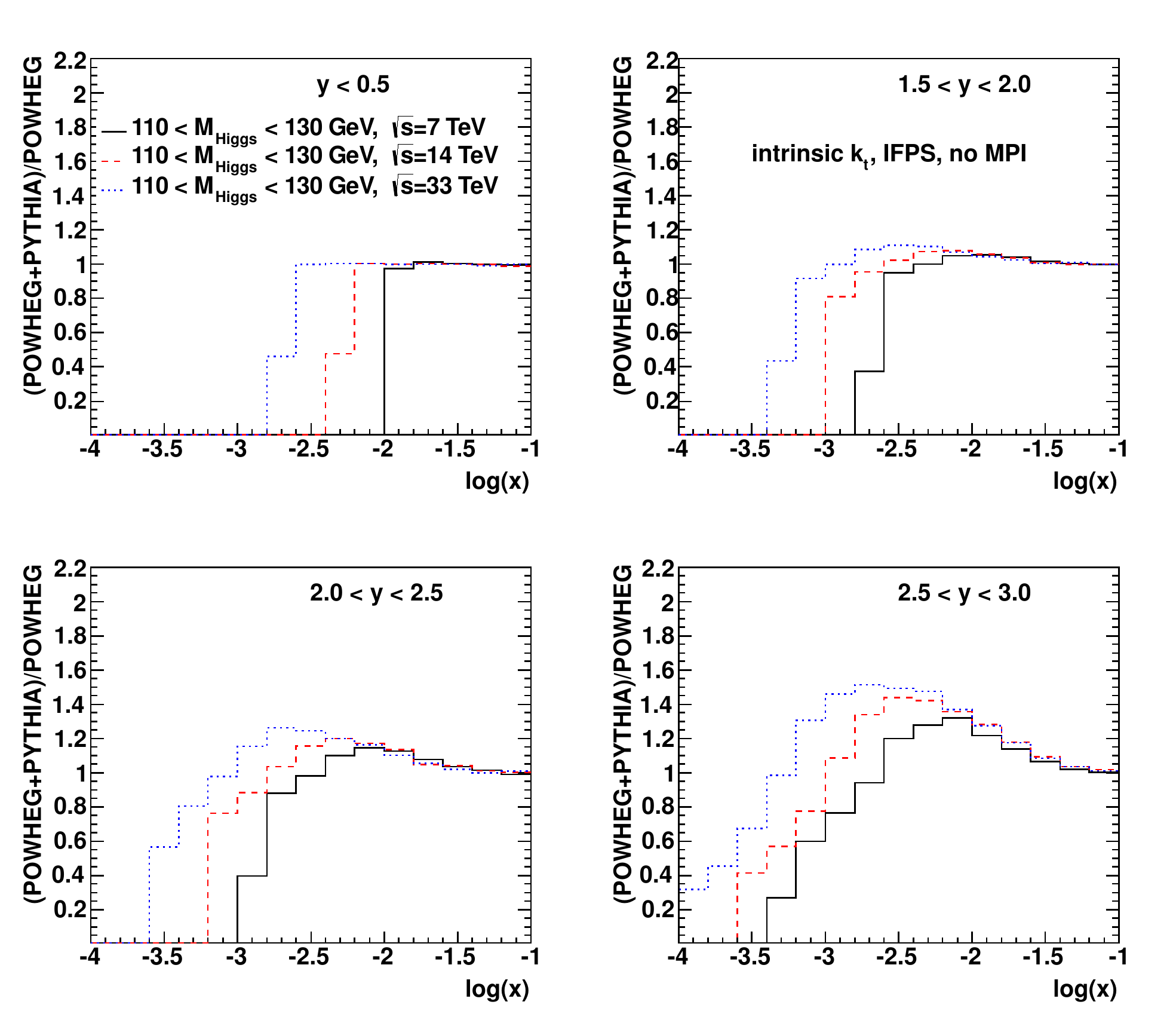}
\caption{\it  Ratio of the cross sections obtained with \protect\powheg\   after  and 
before  inclusion of initial + final state parton shower and intrinsic $k_t$ for  Higgs production at 
different  energies:  $\sqrt{s}=7, 14, 33 $~TeV.
} 
\label{fig:fig6}
\end{figure}

 
 Fig.~\ref{fig:fig5}   
 summarizes the results in Figs.~\ref{fig:fig1}-\ref{fig:fig4}  for     
   the ratio of the cross section   obtained by  \powheg\   after  inclusion of 
     parton showering 
   to the cross section  before  parton showering, plotted for different processes. 
   In Fig.~\ref{fig:fig6} we plot this ratio  
      for Higgs boson production at  different $\sqrt{s}$ energies of  $7, 14$ and $33$~GeV.
      
The longitudinal momentum  shifts  from parton showering  computed in this section 
  measure 
effects  from   QCD  radiation beyond  perturbative  fixed-order  
calculations,  and provide  a  significant 
 contribution  to the correction  factors    in Sec.~II.   
 They affect   initial state showers and    
  need to be consistently  taken into account   
   in calculations which are used to determine parton density 
   functions.    
   The origin of the  kinematical  shifts  lies with the approximation of 
   collinearity~\cite{coki-1209} 
 on the partonic states to which the branching algorithms    describing showers 
 are applied. Although for explicit calculations 
 we have used  a  particular  NLO-shower  matching 
 scheme (\powheg),  the effect    is common  to any calculation matching NLO 
 with collinear showers. 
   In calculations  using integrated parton density functions 
   the correction factors  
  studied in this paper  
    have to be applied after the  evaluation of the cross section (and, as 
    remarked on earlier,   this 
  may   induce systematic inconsistencies if  these  corrections  are 
    not  taken into account properly).      
  On the other hand,   this is avoided  in approaches 
   using transverse momentum dependent  
    PDFs~\cite{jccbook,avsar11,unint09,muld-rog-rev,martin}  
     from the beginning 
     (TMDs or uPDFs),  
     as is done for example in the \cascade\ event generator~\cite{Jung:2010si}.

\section{Conclusions}

Theoretical predictions for high-energy collider processes containing hadronic 
jets require supplementing finite-order 
 perturbative calculations  with  parton showering and 
   nonperturbative  corrections. 
   In this paper we have studied methods to treat 
   parton showering and 
   nonperturbative  corrections   in the context of matched 
   NLO-shower   event generators. 

We have pointed out potential  inconsistencies in current approaches 
which  on the one hand  apply NP correction factors  from 
 leading-order Monte Carlo generators 
to NLO parton-level predictions and on the other hand  fail to include  showering 
corrections. We have proposed methods to address these deficiencies 
by using  consistently  available NLO Monte Carlo tools.   We have shown that the 
differences   in the predictions  for jet cross sections  
induced   by the   modified  approach we propose 
are significant  in  regions of phase space   
which are explored with hard probes    for the first time at the LHC.  In particular, 
the nonperturbative correction factor $K^{NP}$ introduced in Sec.~II 
gives non-negligible differences at low to intermediate jet $p_T$, and 
  the showering  correction factor  $K^{PS}$  of  Sec.~II  
  gives  significant effects  over the whole $p_T$  range  and is largest at 
  large  jet  rapidities $y$.   
  
  Because of this     $y$ and $p_T$ dependence,   
    taking properly into account    NP and showering correction factors     
  changes  the  shape  of jet distributions and  affects significantly   the comparison 
  of theory predictions with experimental data. The numerical results we have presented 
  show effects as large as   50 percent in 
     regions  of $y$ and $p_T$ phase space   relevant to jet 
 measurements at the LHC.   The 
 showering correction factor  $K^{PS}$ in particular can affect the determination of 
 parton distribution functions from fits to experimental   data sets  comprising 
    inclusive  jet  measurements.  

We have  investigated in closer  detail the sources 
of the showering correction  from  initial state   and  final state 
effects. We 
have   observed  that the main initial state  showering effect comes 
from  kinematical  shifts  in   longitudinal momentum   distributions~\cite{coki-1209}   
due to combining collinearity approximations   with  the Monte Carlo 
implementation of  energy-momentum conservation constraints. 
We have examined the   longitudinal   shifts for  specific   processes  in Sec.~III. 
 This effect is largest  for inclusive    jets and 
  $b$-flavor  jets  at the LHC     in the  higher   rapidity bins.   
We have extended the study of  longitudinal  shifts~\cite{coki-1209} 
  to  the case of  Drell-Yan pair production 
by analyzing  the Drell-Yan mass region     $16 < m_{DY} < 166$~GeV  
and found that the shifts  are non-negligible for  
Drell-Yan  production at forward rapidities $ y  \geq 2$. 
 We have also examined the  case of Higgs boson production  for 
 $110 < m_{Higgs} < 130$~GeV 
 and  found that  the  shifts are  non-negligible at large rapidities  at  
 $\sqrt{s}=7$~GeV,  and   become more and more important  at higher
 centre-of-mass energies. 
 
 It will be   interesting  to study the impact 
 of  the   effects discussed in this work    
 on  phenomenological analyses of  LHC final states 
  involving hadronic jets.   We expect  these effects  to   also  
  influence  determinations of parton distributions. 
 Longitudinal momentum shifts  can be avoided in 
formulations  that keep track of non-collinear  (i.e., transverse and/or 
anti-collinear)  momentum components  from the beginning 
using unintegrated  initial state  distributions~\cite{avsar11,unint09}, also 
at parton shower level~\cite{Jung:2010si,jadach09}.   
It will be interesting to  investigate 
 to what extent   this can be exploited to  construct     
approaches  in which nonperturbative contributions such as 
multiple parton interactions, finite transverse momenta,  hadronization  are consistently 
 incorporated in   parton  branching   event generators.   
 
\section*{Acknowledgements}
We are grateful to Torbj\"orn Sj\"ostrand for many discussions concerning hadronisation corrections and multi-parton interactions in \pythia . We are also grateful for many discussions and clarifications on \powheg\  to Simone Alioli.

\end{document}